\documentclass[
reprint,
superscriptaddress,
amsmath,amssymb,
aps,
prb,
]{revtex4-2}

\usepackage{graphicx}
\usepackage{dcolumn}
\usepackage{bm}
\usepackage{changes}
\usepackage{braket}
\usepackage{hyperref}

\begin{document}
\title{Interference-Induced Suppression of Doublon Transport and Prethermalization in the Extended Bose-Hubbard Model}

\author{Zhen-Ting Bao}
\affiliation{Beijing National Laboratory for Condensed Matter Physics, Institute of Physics, Chinese Academy of Sciences, Beijing 100190, China}
\affiliation{School of Physical Sciences, University of Chinese Academy of Sciences, Beijing 100049, China}

\author{Kai Xu}
\email{kaixu@iphy.ac.cn}
\affiliation{Beijing National Laboratory for Condensed Matter Physics, Institute of Physics, Chinese Academy of Sciences, Beijing 100190, China}
\affiliation{School of Physical Sciences, University of Chinese Academy of Sciences, Beijing 100049, China}
\affiliation{Beijing Key Laboratory of Fault-Tolerant Quantum Computing, Beijing Academy of Quantum Information Sciences, Beijing 100193, China}
\affiliation{Hefei National Laboratory, Hefei 230088, China}
\affiliation{Songshan Lake Materials Laboratory, Dongguan, Guangdong 523808, China}

\author{Heng Fan}
\email{hfan@iphy.ac.cn}
\affiliation{Beijing National Laboratory for Condensed Matter Physics, Institute of Physics, Chinese Academy of Sciences, Beijing 100190, China}
\affiliation{School of Physical Sciences, University of Chinese Academy of Sciences, Beijing 100049, China}
\affiliation{Beijing Key Laboratory of Fault-Tolerant Quantum Computing, Beijing Academy of Quantum Information Sciences, Beijing 100193, China}
\affiliation{Hefei National Laboratory, Hefei 230088, China}
\affiliation{Songshan Lake Materials Laboratory, Dongguan, Guangdong 523808, China}

\begin{abstract}

The coherent mobility of doublons, arising from second-order virtual dissociation-recombination processes, fundamentally limits their use as information carriers in the strongly interacting Bose-Hubbard model. We propose a disorder-free suppression mechanism by introducing an optimized nearest-neighbor pair-hopping term that destructively interferes with the dominant virtual hopping channel. Using the third-order Schrieffer-Wolff transformation (SWT), we derive an analytical optimal condition that accounts for lattice geometry corrections. Numerical simulations demonstrate that this optimized scheme achieves near-complete dynamical arrest and entanglement preservation in one-dimensional (1D) chains, while in two-dimensional (2D) square lattices, it significantly suppresses ballistic spreading yet permits a slow residual expansion. Furthermore, in the many-body regime, finite-size scaling analysis identifies the observed long-lived density-wave order as a prethermal plateau emerging from the dramatic separation of microscopic and thermalization timescales.

\end{abstract}
\maketitle

\section{Introduction}

The Bose-Hubbard (BH) model stands as a cornerstone of condensed matter physics, providing a fundamental framework for understanding strongly correlated bosonic matter \cite{fisher1989boson, jaksch1998cold, greiner2002quantum}. From granular superconductors \cite{efetov1980phase, orr1986global, beloborodov2007granular} and Josephson junction arrays \cite{bradley1984quantum, fazio1991charge, fisher1989boson, rosario2001quantum, trugenberger2023gaugge} to modern quantum simulators \cite{georgescu2014quantum, altman2021quantum, grunhaupt2019granular, kamenov2020granular}, the BH model captures the essence of the competition between hopping and on-site interactions, epitomized by the superfluid-Mott insulator transition. Within the strongly interacting regime, stable bound pairs of particles, known as ``doublons'' \cite{winkler2006repulsively, folling2007direct}, emerge as key composite excitations. Due to their stability against dissociation, doublons are regarded as promising candidates for encoding quantum information  and investigating effective pairing mechanisms in lattice systems \cite{duan2003controlling, folling2007direct, winkler2006repulsively}.

However, the utility of doublons is compromised by an intrinsic residual mobility. Even in the limit of strong interactions, doublons are not strictly stationary but possess a mobility arising from second-order virtual processes  \cite{winkler2006repulsively, duan2003controlling, kuklov2003counterflow, folling2007direct}. This effective tunneling drives the coherent transport of doublons, leading to inevitable thermalization \cite{deutsch1991quantum, srednicki1994chaos, li2025many} and the loss of stored information. While introducing disorder can induce many-body localization (MBL) \cite{basko2006metal, oganesyan2007localization, de2014scenario, nandkishore2015many, abanin2019colloquium, xu2018emulating} to arrest this transport, such approaches break the translational symmetry of the system. Achieving robust storage in a clean, translationally invariant setting remains a significant challenge.

In this work, we propose a Hamiltonian engineering strategy to suppress this intrinsic transport via coherent quantum interference. By extending the BH model with a nearest-neighbor (NN) pair-hopping term \cite{dutta2015non, wang2013extended, heng2019pair, ahmadi2025pair}, we introduce a transport channel that destructively interferes with the dominant second-order virtual processes. Going beyond heuristic second-order corrections, we employ a third-order SWT \cite{schrieffer1966relation, zhang1988effective, barthel2009magnetism, bravyi2011schrieffer} to rigorously derive the effective dynamics. Our analysis reveals that lattice connectivity imposes a geometric interference factor $\eta$, which necessitates a precise tuning of the control parameters to the effective hopping.

We validate this mechanism through exact numerical simulations, demonstrating that the optimized interference significantly suppresses ballistic spreading \cite{preiss2015strongly} and preserves entanglement. Specifically, this optimization achieves near-complete dynamical arrest in 1D-BH chain, whereas in 2D-BH square lattice, suppression remains substantial but is limited by residual higher-order pathways due to increased coordination. In the many-body regime, we employ finite-size scaling to strictly distinguish the observed dynamics from true MBL or Hilbert space fragmentation (HSF) \cite{sala2020ergodicity, khemani2020localization, moudgalya2022quantum, wang2025exploring}. Our analysis reveals that the persistent density-wave order is inherently prethermal \cite{gring2012relaxation, abanin2017effective, mori2018thermalization, liu2026prethermalization}, originating from the vast disparity between the suppressed effective hopping and the prolonged thermalization timescales.

\section{Model and Method}\label{sec:model}

\begin{figure*}[]
    \centering
    \includegraphics[]{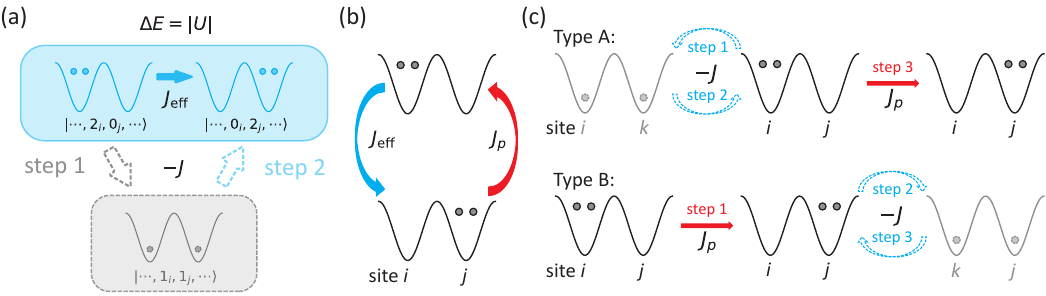}
    \caption{Schematic of doublon dynamics and the suppression mechanism. (a) Second-order tunneling. A doublon tunnels from site $i$ to $j$ via an intermediate state $\ket{\dots, 1_i, 1_j, \dots}$, overcoming the interaction energy gap $\Delta E = \left|U\right|$. This process generates the effective hopping $J_{\text{eff}}$ (blue arrow). (b) Destructive interference. The intrinsic effective amplitude $J_{\text{eff}}$ is canceled by the explicit pair-hopping strength $J_p$ (red arrow) when $J_p \approx -J_{\text{eff}}$. (c) Third-order tunneling. Third-order pathways involving a neighbor $k$. Type A represents virtual processes preceding the pair hop, while Type B shows the converse sequence. Note that $k$ represents all nearest neighbors, including $k=j$ (in Type A) and $k=i$ (in Type B), giving rise to the geometric factor $\eta$.}
    \label{fig:sche}
\end{figure*}

\subsection{Doublon Dynamics in the Standard Bose-Hubbard Model}
We start by considering the standard BH model describing interacting bosons on a $d$-dimensional lattice. Working within the particle-number conserving subspace, the chemical potential term $-\mu\sum_{i}\hat{n}_i$ contributes only a global energy shift and can therefore be omitted. The resulting Hamiltonian (with $\hbar=1$) reads
\begin{align}
    \hat{H}_{\mathrm{BH}}=-J\sum_{\braket{i,j}}\hat{a}^\dagger_i\hat{a}_{j}+\frac{U}{2}\sum_{i} \hat{n}_i\left(\hat{n}_i-1\right),
\end{align}
where $\hat{a}_{i}^{\dagger}$ ($\hat{a}_{i}$) creates (annihilates) a boson at site $i$, and $\hat{n}_{i}=\hat{a}_{i}^{\dagger}\hat{a}_{i}$ is the number operator. The parameters $J$ and $U$ represent the single-particle hopping amplitude and the on-site interaction strength, respectively. The summation $\braket{i,j}$ runs over NN sites. 

In the strong-interaction regime ($u=\left|U/J\right|\gg 1$), single-particle tunneling is significantly suppressed due to the large on-site interactions \cite{fisher1989boson, greiner2002quantum}. Nevertheless, doublons are not strictly localized. Although the standard Hamiltonian $\hat{H}_{\text{BH}}$ contains no direct pair-tunneling term, a doublon on site $i$ can still move via a second-order perturbative process: it virtually dissociates into two singly occupied sites (i.e., $\ket{\dots, 1_i, 1_j, \dots}$) via a single-particle hop, and subsequently recombines at the neighboring site through a second hop [see Fig.~\ref{fig:sche}(a)].

This virtual dissociation--recombination mechanism leads to an effective NN pair-hopping amplitude $J_{\text{eff}} = 2J^2/U$ for doublons \cite{winkler2006repulsively, folling2007direct}. This amplitude, though suppressed by $U$, remains the dominant transport channel, enabling coherent propagation of doublons and ultimately leading to thermalization of the system. Consequently, even in the limit of strong interactions, quantum information encoded in the doublon positions is degraded over time scales proportional to $U/J^2$. This intrinsic residual mobility poses a fundamental challenge for quantum state storage and manipulation, motivating the search for mechanisms to suppress these effective tunneling processes.

\subsection{The Extended Model with Pair Hopping}

To suppress the intrinsic spreading arising from second-order virtual processes, we introduce an explicit control mechanism by extending the standard model with a NN pair-hopping term. Unless otherwise specified, we refer to this interaction simply as ``pair hopping'' throughout the remainder of this work. The resulting Extended Bose-Hubbard (EBH) model Hamiltonian is
\begin{align}
    \hat{H}_{\text{EBH}} = \hat{H}_{\text{BH}} + \hat{H}_{p},
\end{align}
where the pair-hopping term $\hat{H}_{p}$ takes the form
\begin{align}
    \hat{H}_{p} = \frac{J_{p}}{2} \sum_{\braket{i,j}} \hat{p}_{i}^{\dagger}\hat{p}_{j}. 
\end{align}
Here, the operator $\hat{p}_{i} = \hat{a}_{i}^{2}$ annihilates two particles simultaneously at site $i$, so that $\hat{p}_{i}^{\dagger}\hat{p}_{j}$ represents the direct tunneling of a doublon from site $j$ to site $i$. The parameter $J_{p}$ serves as the corresponding hopping strength.

Specifically, the direct pair-hopping amplitude $J_p$ opens a coherent channel that interferes with the virtually induced amplitude $J_{\text{eff}}$ [see Fig.~\ref{fig:sche}(b)]. By tuning $J_p$ to satisfy the destructive interference condition $J_p \approx -J_{\text{eff}}$, the two pathways cancel out. While this cancellation does not strictly arrest all dynamics due to higher-order processes, it eliminates the dominant transport mechanism, thereby resulting in a drastic reduction of doublon mobility. This targeted interference offers a promising mechanism to significantly prolong the storage time of quantum information by overcoming the intrinsic spreading limitations of the standard BHM.

\subsection{Optimal Nearest-Neighbor Pair-Hopping Amplitude}\label{sec:opt_j}

While the condition $J_p \approx -J_{\text{eff}}$ provides a heuristic basis for suppressing spreading, a rigorous derivation requires accounting for higher-order corrections arising from hybrid processes involving both the bare hopping $J$ and the introduced pair hopping $J_p$. Since the pair-hopping term $\hat{H}_p$ enables direct transitions within the doublon subspace, it actively participates in higher-order virtual processes, thereby renormalizing the effective pair-hopping amplitude.

To determine the explicit form of this renormalization, we employ the SWT to block-diagonalize the Hamiltonian up to the third order. The rigorous deduction is detailed in Appendix~\ref{app:third_order}. In contrast to the standard case ($J_p=0$) where the third-order correction strictly vanishes, the inclusion of the pair-hopping term yields a non-vanishing contribution scaling as $\mathcal{O}(J^2 J_p / U^2)$. Physically, this correction arises from two distinct quantum pathways: one sequence where the virtual dissociation-recombination mechanism (induced by the bare hopping $J$) precedes the direct pair hop (Type A), and the converse scenario where the direct pair hop occurs prior to the virtual transport processes (Type B) [see Fig.~\ref{fig:sche}(c)].

Incorporating these contributions, the renormalized effective hopping amplitude $\tilde{J}_{\text{eff}}$ for a doublon between NN sites takes the form
\begin{align}
    \tilde{J}_{\text{eff}} = J_p + \frac{2J^2}{U} - \eta \frac{J^2 J_p}{U^2},
\end{align}
where the first term represents the explicit pair-hopping amplitude, the second term corresponds to the intrinsic second-order hopping generated by the standard BH terms, and the third term denotes the third-order interference correction derived from the aforementioned hybrid processes. The geometric factor $\eta$ accounts for the lattice connectivity. Specifically, for a lattice with coordination number $z$, summing over the available intermediate virtual steps leads to the direct relation $\eta = 2z$. This factor of $2$ arises precisely because the two distinct interference pathways (Type A and Type B) can each proceed via any of the $z$ nearest neighbors.

To achieve maximal suppression of quantum transport, we require the total effective amplitude to vanish, i.e., $\tilde{J}_{\text{eff}} = 0$. Solving this equation yields the optimal pair-hopping strength $J_{p}^{\text{opt}}$.

For a 1D-BH chain, the non-boundary sites interact with two nearest neighbors ($z=2$), leading to a geometric factor of $\eta = 4$. Substituting this into the vanishing condition yields the optimal parameter
\begin{align}
    J_{p}^{\text{opt, 1D}} = \frac{2J^2 U}{4J^2 - U^2}.\label{eq:opt1}
\end{align}

Similarly, in the case of a 2D-BH square lattice, the coordination number is $z=4$, which corresponds to $\eta = 8$. Consequently, the optimal parameter is modified to
\begin{align}
    J_{p}^{\text{opt, 2D}} = \frac{2J^2 U}{8J^2 - U^2}.
\end{align}

It is important to emphasize the physical scope of this result. The condition $\tilde{J}_{\text{eff}} = 0$ is designed to eliminate the effective pair hopping strictly between NN sites and only up to the third order in the perturbation expansion, leaving transport channels from higher-order virtual processes uncompensated. 
These remaining terms (fourth-order and beyond) not only introduce further corrections to the pair-hopping amplitude but also inherently induce longer-range tunneling events, such as next-nearest-neighbor (NNN) hopping. Given that these residual processes scale with higher powers of the perturbative parameter $1/u$, their amplitudes are significantly weaker than the dominant channel. 

Furthermore, the cancellation principle presented here is theoretically generalizable: these residual long-range couplings could, in principle, be further suppressed by extending the control scheme to include additional auxiliary terms, such as NNN pair hopping. Even without such additional complexity, our results demonstrate that eliminating the leading-order channel is sufficient to result in a strongly suppressed, albeit not strictly frozen, dynamical state.

\begin{figure*}[t]
    \centering
    \includegraphics[]{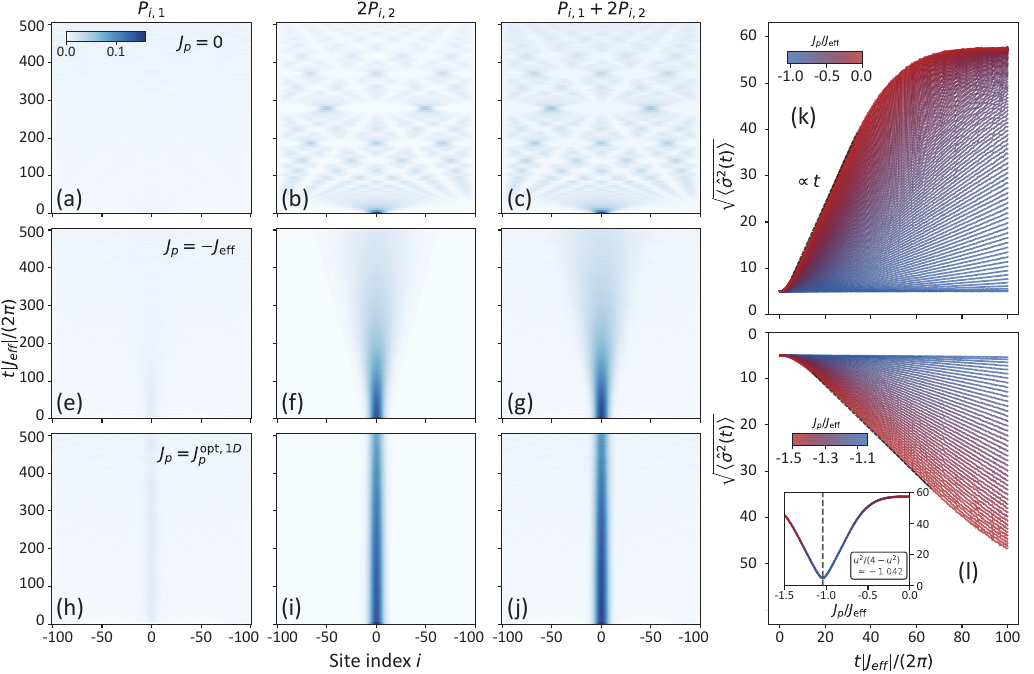}
    \caption{Suppression of doublon transport in a 1D-BH chain with $u=10$. The system is initialized in a Gaussian wave packet ($\sigma=5\sqrt{2}$). (a)-(j) Spatiotemporal evolution of site occupation probabilities up to $t=500 \times 2\pi/\left|J_{\text{eff}}\right|$. Columns from left to right display the single-occupancy probability $P_{i,1}$, the double-occupancy component $2P_{i,2}$, and the total particle density $P_{i,1}+2P_{i,2}$. The three rows correspond to different pair-hopping strengths: (a)-(c) No pair hopping ($J_p=0$); (e)-(g) Heuristic cancellation ($J_p = -J_{\text{eff}}$); and (h)-(j) Optimal cancellation ($J_p = J_p^{\text{opt, 1D}}$). The color bars indicate the magnitude of the corresponding quantities. (k)-(l) Time evolution of the RMSD $\sqrt{\braket{\hat{\sigma}^2(t)}}$ for various ratios of $J_p/J_{\text{eff}}$. The color gradient represents the ratio values. The indicated linear growth regimes ($\propto t$) highlight the ballistic nature of the expansion. The inset in (l) plots the RMSD at $t=100 \times 2\pi/\left|J_{\text{eff}}\right|$ as a function of $J_p/J_{\text{eff}}$, showing a distinct minimum at $\approx -1.042$, consistent with the theoretical optimum derived in Eq.~(\ref{eq:opt1}).}
    \label{fig:1d_diff}
\end{figure*}

\section{Numerical Results}

\begin{figure}[]
    \centering
    \includegraphics[]{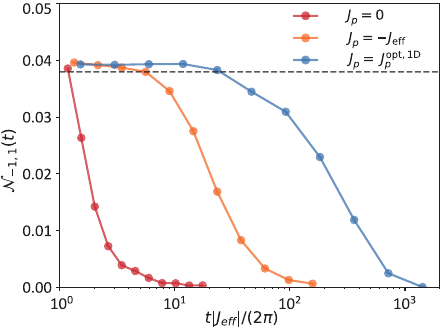}
    \caption{Dynamics of the negativity $\mathcal{N}_{-1,1}(t)$ between symmetric sites $i=-1$ and $j=1$. The time axis is plotted on a logarithmic scale. The system is initialized in a Gaussian wave packet ($\sigma=2\sqrt{2}$). The curves compare the decay of entanglement under three conditions. The dashed horizontal line serves as a visual guide to highlight the significantly prolonged preservation of entanglement in the optimal amplitude.}
    \label{fig:1d_ent}
\end{figure}

To validate the theoretical predictions, we perform exact numerical simulations on finite-size lattices. This approach relies on the conservation of total particle number to strictly truncate the Hilbert space. Unitary time evolution is primarily implemented using Krylov subspace projection methods \cite{park1986unitary, saad2003iterative}. Exclusively for extending our finite-size scaling analysis to larger many-body systems, we employ the time-dependent variational principle (TDVP) algorithm \cite{haegeman2011time, koffel2012entanglement, goto2019performance}. Details of the numerical simulations are provided in Appendix~\ref{app:sim_details}. We then focus on the quench dynamics of two distinct initial states: single-doublon Gaussian wave packets and many-body density-wave orders. The primary goals of our simulations are to quantitatively assess the suppression of doublon transport, the preservation of quantum entanglement, and the emergence and properties of prethermal plateaus under the optimized pair-hopping amplitude.

\subsection{One-Dimensional Dynamics}
The spreading of a single doublon in an empty lattice background is first examined. This provides a clean probe of the effective hopping amplitudes derived in Sec.~\ref{sec:model}. The system is initialized in a Gaussian wave packet centered at site $i_c$, a frequently-used theoretical tool \cite{tutunnikov2023coherent, zhang2024two, zatelli2020scattering, ahmadi2025pair} specifically employed because its well-defined momentum distribution significantly weakens density oscillations, thereby facilitating a clean and rigorous extraction of fundamental transport properties. The Gaussian wave packet reads
\begin{align}
    \ket{\Psi\left(0\right)} \propto \sum_{i} \exp \left[-\frac{d^2\left(i,i_c\right)}{2\sigma^2}\right] \hat{p}_i^\dagger \ket{\mathbf{0}},
\end{align}
where $\sigma$ denotes the width of the Gaussian wave packet, $d(i,i_c)$ is the Euclidean distance between site $i$ and the center $i_c$, and $\ket{\mathbf{0}}$ represents the global vacuum state with no particles on any lattice site. 

Numerical simulations are performed on a 1D-BH chain with $N=201$ sites (from $i=-100$ to $100$), with the interaction strength set to $u=10$ and $J/\left(2\pi\right)=5~\text{MHz}$ \cite{parameter_remark}. The initial state is a Gaussian wave packet with width $\sigma=5\sqrt{2}$ centered at the origin ($i_c=0$). The system is evolved up to a time $t=500 \times 2\pi/\left|J_{\text{eff}}\right|$, and we compare three distinct transport cases: (i) the standard BH model without pair hopping ($J_p=0$), (ii) the heuristic cancellation condition ($J_p = -J_{\text{eff}}$), and (iii) the analytically optimized condition ($J_p = J_p^{\text{opt, 1D}}$).

To characterize the microscopic distribution, we compute the time-dependent probability $P_{i,n}(t)=\mathrm{Tr}[\ket{n}_i\bra{n}_i\hat{\rho}(t)]$ for $n=0,1,2$, describing the particle occupation at site $i$. Here, $\hat{\rho}(t)$ denotes the density matrix of the system at time $t$. The resulting spatiotemporal evolution of the particle density is shown in Fig.~\ref{fig:1d_diff}(a)-(j). In the standard case (top row), the doublon wave packet undergoes ballistic expansion driven by the second-order effective hopping $J_{\text{eff}}$. This effective transport is mediated by virtual dissociation processes, evidenced by the faint yet discernible signatures of single-particle occupation ($P_{i,1}$, left column) accompanying the spreading. When the pair hopping is tuned to $J_p = -J_{\text{eff}}$ (middle row), the expansion is noticeably slowed; however, a residual ``light cone'' remains visible, indicating incomplete cancellation. In contrast, under the optimal condition $J_p = J_p^{\text{opt, 1D}}$ (bottom row), the wave packet exhibits strong dynamical confinement around the center site within the observation window. The stability of the doublon distribution in this regime confirms that the optimized interference effectively inhibits the dominant dissociation--recombination pathways and, as a result, suppresses the leading-order transport.

This suppression is quantitatively substantiated by calculating the mean-square displacement (MSD) of the wave packet. To rigorously isolate the dynamics of the doublons from the dissociated single-particle background, we define the MSD exclusively in terms of the doublon probability distribution:
\begin{align}
    \braket{\hat{\sigma}^2\left(t\right)} = \frac{1}{\sum_{i} P_{i,2}}\sum_{i} P_{i,2} d^2\left(i,i_c\right).
\end{align}

As illustrated in Fig.~\ref{fig:1d_diff}(k)-(l), the spreading dynamics is highly sensitive to the ratio $J_p/J_{\text{eff}}$. While the root-mean-square displacement (RMSD) $\sqrt{\braket{\hat{\sigma}^2\left(t\right)}}$ exhibits characteristic ballistic expansion characterized by approximately linear growth ($\propto t$) over a certain dynamical interval, this spreading is arrested most effectively near the theoretical prediction. The minimum is found at $J_p/J_{\text{eff}} \approx -1.042$, which agrees perfectly with the theoretical value $J_p^{\text{opt,1D}}/J_{\text{eff}}=u^2/(4-u^2) \approx -1.0417$ for $u=10$. This confirms that the third-order correction is essential for precise control in the strong-coupling regime in the 1D-BH chain.

Beyond particle transport, we probe the coherence properties of the system by monitoring the quantum entanglement between the two sites symmetric to the center ($i=-1$ and $j=1$). This is quantified using the negativity \cite{peres1996separability, vidal2002computable}
\begin{align}
    \mathcal{N}_{i,j}(t) = \frac{1}{2} \left( ||\rho_{i,j}^{T_i}(t)||_1 - 1 \right),
\end{align}
 where $\rho_{i,j}(t)$ is the reduced density matrix of sites $i$ and $j$, the superscript $T_i$ denotes the partial transpose with respect to the subspace of site $i$, and $||\cdot||_1$ represents the trace norm. As shown in Fig.~\ref{fig:1d_ent}, for an initial Gaussian wave packet with width $\sigma=2\sqrt{2}$, the negativity in the standard case (red curve) decays rapidly to zero as the doublon delocalizes over the chain. By contrast, under the optimal pair-hopping strength (blue curve), the negativity exhibits a long-lived plateau, indicating that by suppressing the spatial delocalization, our scheme effectively prolongs the lifetime of the quantum correlations in the local basis, protecting them from the rapid decay driven by wave-packet spreading.

\begin{figure}[t]
    \centering
    \includegraphics[]{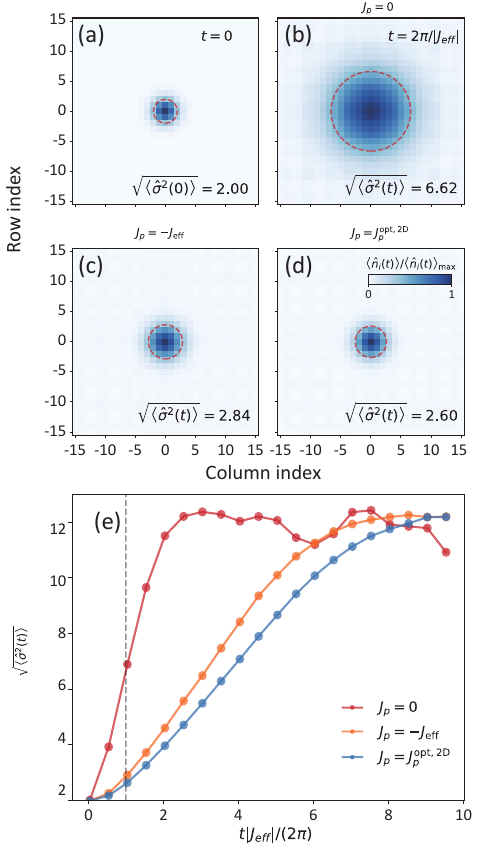}
    \caption{Suppression of doublon transport in a 2D-BH square lattice. (a) The initial particle density distribution of a Gaussian wave packet at $t=0$ with an initial width $\sqrt{\langle\hat{\sigma}^2\left(0\right)\rangle} = 2.00$. (b)-(d) Snapshots of the density distribution at $t=2\pi/\left|J_{\text{eff}}\right|$ for $J_p=0$, $-J_{\text{eff}}$, and $J_p^{\text{opt, 1D}}$, respectively. The values at the bottom right of each panel indicate the corresponding RMSD. The color bar represents the normalized particle density. (e) Time evolution of the RMSD for the three cases. The vertical dashed line indicates the time $t=2\pi/\left|J_{\text{eff}}\right|$ corresponding to the snapshots (b)-(d).} 
    \label{fig:2d_transport}
\end{figure}

\begin{figure*}[ht]
    \centering
    \includegraphics[]{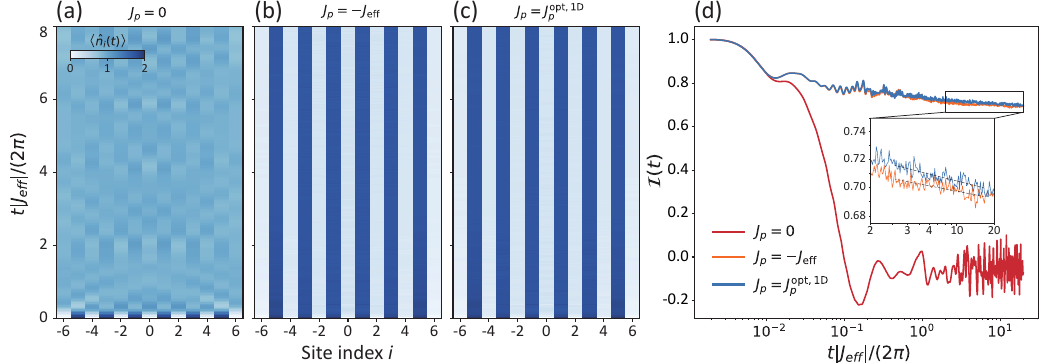}
    \caption{Many-body dynamics of a DW state in a 1D-BH chain with $u=10$. The system is initialized in a DW state $|\Psi_0\rangle = |0,2,0,2,\dots,0\rangle$ with $M=6$ doublons on $N=13$ sites. (a)-(c) Spatiotemporal evolution of the local particle density $\langle \hat{n}_i(t) \rangle$ for $J_p=0$, $-J_{\text{eff}}$, and $J_p^{\text{opt, 1D}}$, respectively. (d) Time evolution of the density imbalance $\mathcal{I}(t)$. The red, orange, and blue curves correspond to the cases in (a), (b) and (c), respectively. The inset shows the decay of $\mathcal{I}(t)$ on a logarithmic scale for the controlled cases in (b) and (c). The dashed lines indicate power-law fits $\mathcal{I}(t) \propto t^{-\beta}$, yielding a vanishingly small exponent $\beta \approx 0.01$ for both scenarios (b) and (c).}
    \label{fig:mb_dynamics}
\end{figure*}

\subsection{Two-Dimensional Dynamics}

The investigation is now extended to the 2D-BH square lattice to verify the universality of the suppression mechanism. As derived in Sec.~\ref{sec:opt_j}, the increased coordination number ($z=4$) in the square lattice modifies the geometric interference factor to $\eta=8$. Consequently, the optimal cancellation condition for the third-order effective hopping shifts to $J_p^{\text{opt, 2D}} = 2J^2 U / (8J^2 - U^2)$.

To visualize the expansion dynamics, we initialize the system with a Gaussian wave packet centered at the middle of a $31 \times 31$ lattice with $961$ sites in total. The parameters are set to match the strong interaction regime used in the 1D case ($u=10$). Fig.~\ref{fig:2d_transport}(a)-(d) displays the snapshots of the particle density distribution at time $t=0$ and $t=2\pi/\left|J_{\text{eff}}\right|$.

In the absence of pair hopping [Fig.~\ref{fig:2d_transport}(b)], the doublon wave packet undergoes significant expansion, driven by the second-order hopping $J_{\text{eff}}$. By setting $J_p = -J_{\text{eff}}$ [Fig.~\ref{fig:2d_transport}(c)], the spreading is visibly suppressed, maintaining a tighter distribution. Most notably, under the optimal condition $J_p = J_p^{\text{opt, 2D}}$ [Fig.~\ref{fig:2d_transport}(d)], the wave packet exhibits the slowest expansion among the three cases, demonstrating a significant suppression of the ballistic dynamics despite the residual spreading.

This behavior is also quantified using the RMSD. As shown in Fig.~\ref{fig:2d_transport}(e), the RMSD for the uncontrolled case (red curve) grows rapidly. Conversely, the optimized case (blue curve) significantly arrests this growth, yielding a RMSD of $\approx 2.60$ at $t=2\pi/\left|J_{\text{eff}}\right|$, compared to $\approx 6.62$ for the uncontrolled case.

We note that while suppression remains substantial, the residual transport in 2D-BH square lattice is more pronounced than in 1D cases (where the wave packet was nearly frozen). This difference arises from the higher lattice connectivity in 2D situations. The increased number of neighbors opens up significantly more pathways for fourth- and higher-order virtual processes, which are not canceled by our third-order optimization scheme. Nevertheless, the results confirm that the interference mechanism remains robust and effective in higher spatial dimensions.

\subsection{Many-Body Dynamics and Prethermalization}

Having established the efficacy of the optimized pair hopping in suppressing single-doublon transport, we next examine its validity in the many-body regime, where interactions between multiple doublons could destabilize the confinement mechanism---a crucial step for assessing whether the interference-induced suppression of transport survives at finite densities.

For our simulations, we consider a 1D-BH chain with $N=13$ lattice sites and $M=\lfloor N/2 \rfloor=6$ doublons. The system is initialized in a density wave (DW) state \cite{trotzky2012probing}, formed by an array of doublons separated by empty sites:
\begin{equation}
    |\Psi(0)\rangle = |0, 2, 0, 2, \dots, 0, 2, 0\rangle.
    \label{eq:cdw_state}
\end{equation}

We then monitor the time evolution of the local particle density $\langle \hat{n}_i(t) \rangle$ and quantify the preservation of the crystalline order using the density imbalance \cite{schreiber2015observation}:
\begin{equation}
    \mathcal{I}(t) = \frac{1}{2M} \sum_{i=-M}^{M} (-1)^{i+1} \langle \hat{n}_i(t) \rangle.
    \label{eq:imbalance}
\end{equation}

For the perfect DW state, $\mathcal{I}(0) = 1$, while $\mathcal{I} \sim 0$ signifies thermalization to a uniform density distribution. We specifically rely on this observable because, unlike entanglement-based measures that demand resource-intensive quantum state tomography (QST) \cite{brydges2019probing, kaufman2016quantum, huang2020predicting}, it can be directly extracted from single-site local measurements, making our predictions readily verifiable on near-term quantum simulators \cite{xu2018emulating, guo2021observation, li2025many, liu2026prethermalization}. (Supplementary calculations of the bipartite entanglement entropy are provided in Appendix~\ref{app:entropy}.)

Fig.~\ref{fig:mb_dynamics}(a)-(c) illustrates the stark contrast in dynamics. In the standard BH model ($J_p=0$, see Fig.~\ref{fig:mb_dynamics}(a)), the DW pattern melts rapidly as doublons diffuse and collide, leading to a thermalized state. Conversely, under the optimal pair-hopping condition ($J_p = J_p^{\text{opt, 1D}}$, see Fig.~\ref{fig:mb_dynamics}(c)), the density profile remains remarkably stable over the simulated time scale ($t \sim 8 \times 2\pi/\left|J_{\text{eff}}\right|$), exhibiting a dynamical arrest of the initial order.

This suppression is quantitatively captured by the imbalance dynamics shown in Fig.~\ref{fig:mb_dynamics}(d). While the uncontrolled system (red curve) exhibits a fast decay to zero, the optimally controlled system (blue curve) maintains a long-lived plateau with $\mathcal{I}(t) \approx 0.7$.
Importantly, this state is not strictly static. As shown in the inset of Fig.~\ref{fig:mb_dynamics}(d), the imbalance exhibits a slow power-law decay $\mathcal{I}(t) \propto t^{-\beta}$ with a very small exponent $\beta \approx 0.01$. This slow relaxation dynamics is a hallmark of prethermalization. The system is trapped in a metastable non-equilibrium state due to the suppression of the leading-order kinetic terms, postponing thermalization to timescales governed by higher-order perturbative processes.

\begin{figure*}[ht]
    \centering
    \includegraphics[]{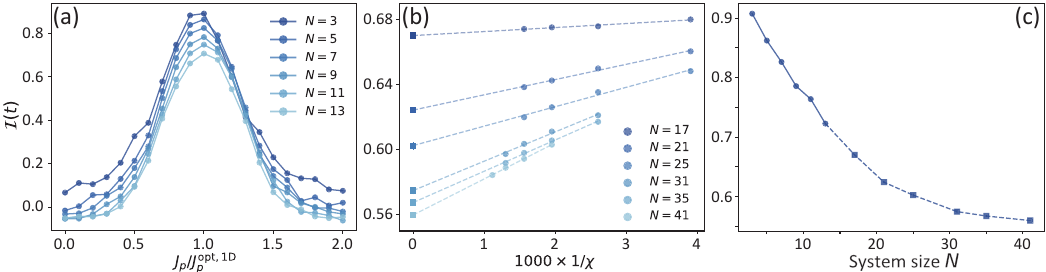}
    \caption{Finite-size scaling of the long-time imbalance. (a) The imbalance $\mathcal{I}(t)$ is averaged over a late-time window around $t=10 \times 2\pi/\left|J_{\text{eff}}\right|$ and plotted as a function of the pair-hopping strength ratio $J_p / J_p^{\text{opt, 1D}}$. Different curves correspond to different system sizes ranging from $N=3$ to $13$. (b) The imbalance evaluated at $t=2 \times 2\pi/\left|J_{\text{eff}}\right|$ for larger system sizes ($N > 13$), calculated using the TDVP algorithm with various bond dimensions $\chi$. The dashed lines represent linear fits used for extrapolation to the infinite bond dimension limit, with the predicted values denoted by square markers. Error bars indicate the fitting uncertainty but are negligibly small and largely obscured by the markers. (c) The imbalance at $t=2 \times 2\pi/\left|J_{\text{eff}}\right|$ plotted as a function of the system size $N$. Circular markers represent exact numerical results obtained via the Krylov subspace projection methods for smaller systems, while square markers represent the TDVP-extrapolated predictions, corresponding directly to the square markers in panel (b).}
    \label{fig:mb_fss}
\end{figure*}

To elucidate the nature of this dynamical arrest---specifically, to distinguish between true MBL or HSF and prethermalization---we perform a finite-size scaling analysis. Fig.~\ref{fig:mb_fss} displays the finite-size scaling analysis of the long-time imbalance. For smaller system sizes ranging from $N=3$ to $13$, calculated via Krylov subspace methods, the imbalance as a function of the control parameter ratio $J_p/J_p^{\text{opt, 1D}}$ is shown in Fig.~\ref{fig:mb_fss}(a). A key feature is observable here: the peak of the imbalance consistently aligns with the theoretical optimum $J_p/J_p^{\text{opt, 1D}} \approx 1$ for all evaluated small sizes, validating that the local interference mechanism remains dominant in the many-body setting.

To further elucidate the nature of this dynamical arrest at macroscopic scales, we extend our scaling analysis to larger systems up to $N=41$ using the TDVP algorithm. To ensure the highest accuracy of the tensor network calculations, the imbalance at a representative plateau time is simulated using various bond dimensions $\chi$ and rigorously extrapolated to the infinite bond dimension limit ($1/\chi \to 0$), as shown in Fig.~\ref{fig:mb_fss}(b). By combining the exact numerical results for smaller systems with the TDVP-extrapolated predictions, Fig.~\ref{fig:mb_fss}(c) illustrates the scaling behavior of the long-time density imbalance with respect to the system size $N$. Notably, while the imbalance experiences an initial decrease at smaller sizes, the curve gradually levels off and approaches a significant non-zero value as $N$ increases further. This specific leveling-off scaling behavior holds decisive physical significance: it explicitly rules out the possibility that the observed dynamical arrest is merely an artifact of finite-size effects. Instead, it conclusively demonstrates that the prethermal plateau, originating from the vast disparity between the suppressed effective hopping and the prolonged thermalization timescales, remains fundamentally robust in the macroscopic limit.

\section{Discussion and Conclusion}

In this work, we have presented a comprehensive strategy to suppress the intrinsic mobility of doublons in the strongly interacting BH model. By engineering a destructive interference between the virtually induced second-order hopping and an explicit pair-hopping term, we have demonstrated the capability to arrest the dominant transport channels in a clean, disorder-free system. We note that an analogous dynamical arrest due to competing hopping pathways has recently been observed in ferromagnetic systems \cite{paul2025anomalous}, suggesting that the suppression of bound state transport via destructive interference is a robust mechanism across different physical setups.

A central theoretical contribution of our study is the rigorous derivation of the optimal cancellation condition beyond the heuristic level. While a simple cancellation of the second-order amplitude ($J_p \approx -J_{\text{eff}}$) offers qualitative suppression, our third-order SWT analysis reveals that precise control requires accounting for hybrid tunneling processes. We identified a geometric interference factor $\eta$, which quantifies the renormalization of the effective pair hopping due to lattice connectivity. The excellent agreement between our analytical predictions and the numerical minima of wave-packet spreading, most notably in the 1D-BH chain, confirms that these higher-order corrections are non-negligible in the strong-coupling regime.

Our analysis of dimensionality reveals that the efficacy of dynamical arrest is sensitive to lattice geometry. In 1D-BH chain, the optimized control leads to a near-complete freezing of doublon dynamics, whereas 2D-BH square lattice exhibit substantial suppression accompanied by a slow, residual expansion. This distinct behavior stems from the increased coordination number ($z=4$), which significantly multiplies the available pathways for higher-order virtual processes. Unlike the targeted third-order cancellation, these uncompensated contributions introduce further corrections to the pair-hopping amplitude while simultaneously inducing longer-range hoppings. Although the theoretical optimization minimizes the dominant transport channel up to third-order correction, the cumulative effect of these residual pathways prevents the strict localization. Despite this geometric limitation, the drastic reduction in mobility confirms the robustness of the interference mechanism in arresting the leading-order dynamics against lattice complexity.

In the many-body context, our results demonstrate the applicability of few-body coherent control strategies to many-body non-equilibrium regimes. The observation of a long-lived density-wave plateau under optimal control signifies the emergence of prethermalization. Crucially, our extended scaling analysis confirms that this dynamical arrest is not a finite-size artifact but a robust macroscopic phenomenon. Although the system ultimately thermalizes, the destructive interference guarantees a profound timescale separation.Specifically, under the optimal pair-hopping condition ($J_p = J_p^{\text{opt}}$), our perturbative analysis indicates that the uncompensated fourth-order virtual processes dictate a residual hopping amplitude $J_{\text{res}} \sim \mathcal{O}(J^4/U^3)$. This theoretically extends the characteristic time scale of the metastable window to $\tau \propto U^3/J^4$, representing a dramatic enhancement compared to the uncontrolled thermalization timescale $\tau_0 \propto U/J^2$. This ``dynamical arrest'' effectively creates a metastable window that potentially offers beneficial conditions for future quantum information storage and manipulation.

Experimental realization of the BH model has been established on the platform of superconducting circuits \cite{bianchetti2010control, roushan2017spectroscopic, yan2019strongly, ye2019propagation, wang2025observing, ticea2025observation}, particularly arrays of transmon qutrits \cite{koch2007charge}. In a conventional array, the native capacitive coupling between adjacent transmons is dominated by single-particle hopping, typically described by $H_g \propto (a_i + a_i^\dagger)(a_j + a_j^\dagger)$, where $a_i$ ($a_i^\dagger$) is the annihilation (creation) operator for a single excitation in the $i$-th transmon. Consequently, direct pair-exchange processes are inherently forbidden in the bare state basis, as the matrix element $\bra{20}H_g\ket{02}$ strictly vanishes. Here, the state $\ket{02}=\ket{0}\otimes\ket{2}$ specifically denotes a configuration where one transmon is populated in its second excited state while its neighbor remains in the ground state.

However, this fundamental limitation may be overcome using advanced Floquet engineering techniques \cite{eckardt2017colloquium}. By applying a detuned, time-periodic microwave drive, transmons can be adiabatically mapped into Floquet qutrits \cite{nguyen2024programmable}. Within this framework, the target Hilbert space is spanned by the Floquet modes $\ket{u_n\left(t\right)}_{\text{F}}$, which are engineered time-dependent superpositions of the bare energy levels $\ket{n}$ ($n=0,1,2$). Defining the basis as $\ket{u_{mn}\left(t\right)}_{\text{F}} = \ket{u_{m}\left(t\right)}_{\text{F}} \otimes \ket{u_{n}\left(t\right)}_{\text{F}}$ ($n,m=0,1,2$), the effective interactions are governed by the time-averaged matrix elements of the native coupling $H_g$ evaluated in the Floquet basis. Because the Floquet dressing hybridizes the bare states across different excitation manifolds, the time-averaged transition amplitude $\overline{\bra{u_{20}\left(t\right)}_{\text{F}} H_g \ket{u_{02}\left(t\right)}_{\text{F}}}$ becomes non-zero, thereby generating a direct pair-hopping channel. Crucially, the effective pair-hopping strength $J_p$ can be tuned in magnitude by controlling the amplitude and detuning of the microwave drives, providing a promising theoretical framework to satisfy the exact destructive interference condition derived in this work. Importantly, considering typical transmon parameters ($J/\left(2\pi\right) \sim 10~\text{MHz}$, $U/\left(2\pi\right) \sim -200~\text{MHz}$ ), the characteristic time scale for the effective pair hopping ($2\pi/\left|J_{\text{eff}}\right| \sim 1\ \mu\text{s}$) is significantly shorter than state-of-the-art device coherence times ($> 100\ \mu\text{s}$), making it plausible to observe the primary interference-induced dynamical arrest within the experimental window.

Nevertheless, it is important to acknowledge that physically synthesizing and precisely calibrating these tailored pair-hopping terms in a large-scale array entails complex engineering challenges. Strong microwave driving can introduce both hardware-specific issues, such as microwave crosstalk, and unwanted non-equilibrium effects, including heating and loss of coherence. While devising a comprehensive blueprint to mitigate these intricate engineering challenges is beyond the scope of the present theoretical study, the high degree of controllability and rapid advancements in the dynamic control of superconducting processors make them a challenging but promising candidate for exploring the predicted dynamical arrest and prethermalization.

Beyond superconducting circuits, ultracold atoms in optical lattices represent the most natural platform for simulating the BH model \cite{jaksch1998cold, rosario2001quantum, bloch2012quantum, dutta2015non}. However, when considering the cross-platform realization of our specific Floquet engineering scheme, a fundamental physical distinction emerges. Unlike superconducting transmons, where the local Fock state $\ket{n}$ corresponds to excitation energy levels that can be freely hybridized via a transverse microwave drive (introducing a driving Hamiltonian $ \hat{H}_{\text{d}}\propto \hat{a} + \hat{a}^\dagger$), the state $\ket{n}$ in cold atom simulators is strictly defined by the number of massive physical atoms. Because isolated atomic systems enforce particle number conservation, the transverse-field driving utilized here is unphysical in that context. Consequently, synthesizing analogous pair-hopping dynamics in atomic optical lattices requires alternative pathways.

Looking forward, this work opens several avenues for future exploration. Theoretically, the SWT employed here can be systematically extended to higher orders; such refined calculations would yield precise corrections for residual hoppings, thereby offering a pathway to further enhance the suppression efficiency even in more complex lattice geometries. Beyond higher-order corrections, the interference principle itself could be generalized to suppress longer-range tunneling by introducing corresponding terms. Furthermore, investigating the interplay between this interference-induced dynamical arrest and weak external disorder could yield new insights into the stability of quantum matter and the transition between prethermal and localized phases. Ultimately, our findings provide a concrete blueprint for Hamiltonian engineering in strongly correlated systems, enabling the control of transport properties and enhanced preservation of quantum states through the precise manipulation of virtual processes.

\begin{acknowledgments}
    We thank the support from the Synergetic Extreme Condition User Facility (SECUF) in Huairou District, Beijing. This work was supported by the National Natural Science Foundation of China (Grants No. U25A6009, No. 92265207, No. T2121001, No. T2322030, No. 12122504, No. 12274142, and No. 12475017), the Innovation Program for Quantum Science and Technology (Grant No. 2021ZD0301800), the Beijing Nova Program (Grant No. 20220484121), and the Ministry of Science and Technology project (Grant No. 2025YFE0217600).

    We also thank Jiajun Yu for fruitful discussions regarding the numerical simulations.
\end{acknowledgments}

\appendix

\section{Perturbative Expansion of the Schrieffer-Wolff Transformation}\label{app:swt}

We briefly review the formalism of the SWT and provide the effective Hamiltonian expansion up to the third order. This allows us to systematically decouple the target subspace from the complementary subspace and identify the leading-order spreading term relevant to the main text. We follow the rigorous formulation described in Ref.~\cite{bravyi2011schrieffer}.

Consider a Hamiltonian $\hat{H} = \hat{H}_0 + \hat{V}$, where $\hat{H}_0$ is the unperturbed diagonal Hamiltonian and $\hat{V}$ is the perturbation. Let $\mathcal{P}$ be the target subspace spanned by a chosen set of eigenvectors of $\hat{H}_0$ and $\mathcal{Q}$ be the complementary subspace orthogonal to $\mathcal{P}$. We denote $\hat{P}$ and $\hat{Q} = 1 - \hat{P}$ as the projectors onto $\mathcal{P}$ and $\mathcal{Q}$, respectively.

The goal of the SWT is to find a unitary transformation $e^{\hat{S}}$ that block-diagonalizes the Hamiltonian $\hat{H}$ to the desired order. The generator $\hat{S}$ is required to be anti-hermitian ($\hat{S}^\dagger = -\hat{S}$) and block-off-diagonal ($\hat{P}\hat{S}\hat{P} = \hat{Q}\hat{S}\hat{Q} = 0$). The effective Hamiltonian acting on the target subspace is then given by
\begin{align}
    \hat{H}_{\text{eff}} = \hat{P} e^{\hat{S}} \hat{H} e^{-\hat{S}} \hat{P}.
\end{align}

To explicitly construct the series, it is convenient to introduce the superoperator formalism. We define the block-diagonal and block-off-diagonal parts of an operator $\hat{X}$ as $\hat{X}_\text{d} = \hat{P}\hat{X}\hat{P} + \hat{Q}\hat{X}\hat{Q}$ and $\hat{X}_{\text{od}} = \hat{P}\hat{X}\hat{Q} + \hat{Q}\hat{X}\hat{P}$. A key ingredient is the superoperator $\mathcal{L}$, which determines the transformation generator by solving the Sylvester equation associated with block-diagonalization:
\begin{align}
    \mathcal{L}\left(\hat{X}\right) = \sum_{i,j} \frac{\bra{i}\hat{X}_{\text{od}}\ket{j}}{E_i - E_j} \ket{i}\bra{j},
\end{align}
where $\ket{i} \in \mathcal{P}$ and $\ket{j} \in \mathcal{Q}$ (or vice versa) are eigenstates of $\hat{H}_0$ with energies $E_i, E_j$.

Expanding the generator $\hat{S} = \sum_{n=1}^{+\infty} \hat{S}_n$ and the effective Hamiltonian $\hat{H}_{\text{eff}} = \sum_{n=0}^{+\infty} \hat{H}_{\text{eff}}^{(n)}$ as a power series with respect to the perturbation $\hat{V}$, the first and second-order terms for the generator are determined recursively as
\begin{align}
    \hat{S}_1 &= \mathcal{L}\left(\hat{V}_{\text{od}}\right), \label{eq:S1_gen} \\
    \hat{S}_2 &= -\mathcal{L}\left(\left[\hat{V}_{\text{d}}, \hat{S}_1\right]\right). \label{eq:S2_gen}
\end{align}

By using these generators, we can obtain the effective Hamiltonian terms. The zeroth- and first-order terms are simply the projection of the original Hamiltonian:
\begin{align}
    \hat{H}_{\text{eff}}^{(0)} &= \hat{P} \hat{H}_0 \hat{P}, \\
    \hat{H}_{\text{eff}}^{(1)} &= \hat{P} \hat{V} \hat{P}.
\end{align}

The second- and third-order corrections, which capture the dominant virtual processes and the interference effects discussed in this work, are given by
\begin{align}
    \hat{H}_{\text{eff}}^{(2)} &= \frac{1}{2} \hat{P} \left[\hat{S}_1, \hat{V}_{\text{od}}\right] \hat{P}, \label{eq:Heff2_gen} \\
    \hat{H}_{\text{eff}}^{(3)} &= \frac{1}{2} \hat{P} \left[\hat{S}_2, \hat{V}_{\text{od}}\right] \hat{P}. \label{eq:Heff3_gen}
\end{align}

These expressions form the basis for calculating the renormalized pair-hopping amplitudes in the specific BH model context.

\section{Derivation of the Second-Order Effective Hopping}
\label{app:second_order}

In this appendix, we apply the perturbative expansion of the SWT presented in Appendix~\ref{app:swt} to the standard BH model to derive the second-order effective hopping amplitude for a doublon. This process corresponds to the evaluation of the second-order term $\hat{H}_{\text{eff}}^{(2)}$.

Here, we consider the dynamics of a single doublon in a vacuum. The unperturbed Hamiltonian is $\hat{H}_0 = \left(U/2\right) \sum_i \hat{n}_i(\hat{n}_i-1)$, and the perturbation is the single-particle hopping $\hat{V} = -J \sum_{\langle i,j \rangle} \hat{a}_i^\dagger \hat{a}_j$. The validity of this perturbative approach is strictly ensured by the strong-interaction condition $u\gg 1$, which treats the single-particle hopping as a small disturbance relative to the on-site interactions. Consequently, we can define the target subspace $P$ spanned by the doublon states $|\psi_{i}\rangle=\left(1/\sqrt{2}\right)\hat{p}_{i}^{\dagger}|\mathbf{0}\rangle$ with unperturbed energy $E_{i}=E_{\text{doub}}=U$, well-separated from the complementary subspace $\mathcal{Q}$, which consists of virtual intermediate states $|\phi_{ij}\rangle = \hat{a}_i^\dagger \hat{a}_j^\dagger |\mathbf{0}\rangle$ ($i\neq j$) with energy $E_{i,j}=E_{\text{virt}} = 0$, representing two particles occupying two different sites $i$ and $j$.

The first-order generator $\hat{S}_1$ is obtained via Eq.~(\ref{eq:S1_gen}). The off-diagonal perturbation $\hat{V}_{\text{od}}$ connects $\mathcal{P}$ and $\mathcal{Q}$ only for pairs of NN sites $\langle i,j \rangle$, the matrix element is $\langle \psi_i | \hat{V}_{\text{od}} | \phi_{ij} \rangle = -\sqrt{2}J$, while for non-NN sites, this contribution vanishes. Utilizing the definition of the superoperator $\mathcal{L}$ and the energy gap $E_i-E_{i,j}=E_{\text{doub}}-E_{\text{virt}} = U$, the relevant non-vanishing matrix elements of the generator are
\begin{align}
    \langle \psi_i | \hat{S}_1 | \phi_{ij} \rangle = \frac{\langle \psi_i | \hat{V}_{\text{od}} | \phi_{ij} \rangle}{U} = -\frac{\sqrt{2}J}{U}.
\end{align}

We now evaluate the second-order effective Hamiltonian $\hat{H}_{\text{eff}}^{(2)} = \left(1/2\right)\hat{P} [\hat{S}_1, \hat{V}_{\text{od}}] \hat{P}$. Due to the strict locality of the operators, non-vanishing off-diagonal matrix elements exist only between NN sites $i$ and $j$. Expanding the commutator for such a pair yields
\begin{align}
    \langle \psi_j | \hat{H}_{\text{eff}}^{(2)} | \psi_i \rangle = \frac{1}{2} \sum_{m \in \mathcal{Q}} \Big( &\langle \psi_j | \hat{S}_1 | m \rangle \langle m | \hat{V}_{\text{od}} | \psi_i \rangle \nonumber \\
    &- \langle \psi_j | \hat{V}_{\text{od}} | m \rangle \langle m | \hat{S}_1 | \psi_i \rangle \Big), \label{eq:B3_split}
\end{align}
where the summation over $m$ survives only for the unique intermediate state $m=|\phi_{ij}\rangle$ that links sites $i$ and $j$. Using the relation $\hat{S}_1^\dagger = -\hat{S}_1$, the first term in the bracket evaluates to $(-\sqrt{2}J/U)(-\sqrt{2}J) = 2J^2/U$, and the second term evaluates to $(\sqrt{2}J)(\sqrt{2}J/U) = -2J^2/U$. Substituting these values into Eq.~(\ref{eq:B3_split}), we obtain the off-diagonal matrix element and effective pair-hopping amplitude:
\begin{equation}
    J_{\text{eff}}\equiv \langle \psi_j | \hat{H}_{\text{eff}}^{(2)} | \psi_i \rangle = \frac{2J^2}{U}.
    \label{eq:Heff2_offdiag}
\end{equation}

Furthermore, strictly constructing the complete second-order effective Hamiltonian serves as the basis for the subsequent third-order calculation. Therefore, we also account for the diagonal matrix elements $\langle \psi_i | \hat{H}_{\text{eff}}^{(2)} | \psi_i \rangle$. These terms, usually referred to as the Lamb shift, arise from virtual processes where a doublon dissociates into a neighbor $j$ and recombines at the original site $i$ (i.e., $| \dots, 2_i, 0_j, \dots \rangle \to | \dots, 1_i, 1_j, \dots \rangle \to | \dots, 2_i, 0_j, \dots \rangle$). Similarly, expanding the commutator for the diagonal case yields
\begin{align}
    \langle \psi_i | \hat{H}_{\text{eff}}^{(2)} | \psi_i \rangle &= \frac{1}{2} \sum_{j \in \text{NN}\left(i\right)} \left( \langle \psi_i | \hat{S}_1 | \phi_{ij} \rangle \langle \phi_{ij} | \hat{V}_{\text{od}} | \psi_i \rangle - \text{h.c.} \right)\nonumber\\
    &= \sum_{j \in \text{NN}\left(i\right)} J_{\text{eff}},
\end{align}
where $\text{NN}\left(i\right)$ is the set of all NN sites of $i$. 

Since each NN site contributes an energy shift identical in magnitude to the hopping amplitude, the total diagonal shift for a doublon at site $i$ depends on the lattice coordination number $z$. 

Summing over all NN pairs $\langle i,j \rangle$, we can express the full second-order effective Hamiltonian in operator form, explicitly including the diagonal Lamb shift terms:
\begin{equation}
    \hat{H}_{\text{eff}}^{(2)} = \frac{1}{2} J_{\text{eff}} \sum_{\langle i,j \rangle} \left[\hat{p}_j^\dagger\hat{p}_i + \hat{n}_i\left(\hat{n}_i-1\right) \right].
    \label{eq:Heff2_full}
\end{equation}

Although these diagonal terms in Eq.~(\ref{eq:Heff2_full}) act as a constant energy shift in a uniform infinite lattice, their inclusion ensures the rigorous completeness of the SWT up to the second order, which lays the necessary groundwork for the derivation of third-order corrections.

\section{Third-Order Correction and Optimal Hopping}
\label{app:third_order}

We now extend the analysis to the third-order correction to the effective amplitude $\tilde{J}_{\text{eff}}$ arising from the inclusion of the pair-hopping term $\hat{H}_p$. Before proceeding to the derivation, we first address the validity of the perturbative treatment. Given that thepair-hopping strength is tuned to cancel the second-order effective hopping, its magnitude scales as $J_p \sim \mathcal{O}(J^2/U)$. This scaling ensures that $\hat{H}_p$ remains a consistent perturbation within the hierarchy of the SWT, thereby preserving the convergence of the series. With this established, we show how the interplay between single-particle hopping and pair hopping leads to the geometric factor $\eta$.

\subsection{Third-Order Effective Hamiltonian}
Under the EBH model, the perturbation is $\hat{V} = \hat{V}_J + \hat{H}_p$, where $\hat{V}_J$ is the single-particle hopping and $\hat{H}_p$ is the pair hopping. In the strong-coupling regime, $\hat{H}_p$ acts as a block-diagonal perturbation within the target subspace $\mathcal{P}$, while $\hat{V}_J$ is block-off-diagonal. Thus, we identify $\hat{Q}\hat{H}_p\hat{Q}=0$, $\hat{V}_{\text{d}} = \hat{H}_p$ and $\hat{V}_{\text{od}} = \hat{V}_J$.

From Eq.~\ref{eq:S2_gen}, the second-order generator $\hat{S}_2$ is
\begin{align}
    \hat{S}_2 &= -\mathcal{L}\left(\left[\hat{H}_p, \hat{S}_1\right]\right)\nonumber\\
    &= -\frac{1}{U}\left(\hat{P}\left[\hat{H}_p,\hat{S}_1\right]\hat{Q}-\text{h.c.}\right).
\end{align}
As shown in Appendix~\ref{app:second_order}, $\hat{S}_1$ connects a doublon state $\ket{\psi_i}$ to a virtual state $\ket{\phi_{ik}} \in \mathcal{Q}$. The commutator $[\hat{H}_p, \hat{S}_1]$ generates terms where a pair hop occurs either before or after the action of $\hat{S}_1$. Using Eq.~\ref{eq:Heff3_gen}, the third-order effective Hamiltonian is
\begin{align}
    \hat{H}_{\text{eff}}^{(3)} &= \frac{1}{2} \hat{P} \left[\hat{S}_2, \hat{V}_{J}\right] \hat{P}. \nonumber\\
    &= -\frac{1}{2U}\left(\hat{H}_p\hat{H}_{\text{eff}}^{\left(2\right)}+\text{h.c.}\right).
    \label{eq:C_Heff3}
\end{align}

\subsection{Path Analysis and Geometric Factor}
We seek to calculate the renormalization of the hopping amplitude between two NN sites $i$ and $j$. It is important to note that third-order perturbative processes can generally induce longer-range couplings, such as NNN hopping. However, such terms constitute a distinct transport channel that corresponds to orthogonal operators in the Hamiltonian. Since the control parameter $J_p$ is strictly a pair-hopping amplitude, it cannot act to cancel these longer-range terms. Therefore, to derive the optimal cancellation condition for the dominant transport channel, we restrict our analysis exclusively to the effective matrix elements $\langle \psi_j | \hat{H}_{\text{eff}}^{(3)} | \psi_i \rangle$ where $i$ and $j$ are nearest neighbors.

By expanding Eq.~(\ref{eq:C_Heff3}), we identify two distinct types of transport pathways, as explicitly illustrated in Fig.~\ref{fig:sche}(c). These pathways describe how the doublon transfers from site $i$ to NN site $j$ with the participation of an intermediate neighbor $k$:

Type A (Virtual hop precedes pair hop): As depicted in the top panel of Fig.~\ref{fig:sche}(c), the doublon at the initial site $i$ first undergoes a virtual hop: a single particle tunnels to a neighbor $k$ and immediately tunnels back to $i$. This loop process leaves the doublon at site $i$ but renormalizes its state. Subsequently, the doublon executes a direct pair hop $J_p$ to the destination site $j$. Note that the neighbor $k$ can be any nearest neighbor of $i$, including $j$.

Type B (Pair hop precedes virtual hop): As depicted in the bottom panel of Fig.~\ref{fig:sche}(c), the sequence is reversed. The doublon first executes a direct pair hop $J_p$ from site $i$ to site $j$. Once at site $j$, it undergoes a virtual hop where a single particle tunnels to a neighbor $k$ and returns to $j$. Similarly, $k$ can be any nearest neighbor of $j$, including $i$.

The total amplitude is obtained by summing over all possible intermediate neighbors $k$ allowed by the lattice geometry. Evaluating the commutators explicitly yields the term as
\begin{align}
    \langle \psi_j | \hat{H}_{\text{eff}}^{(3)} | \psi_i \rangle = -\eta \frac{J^2 J_p}{U^2},
\end{align}
where the geometric factor $\eta$ counts the multiplicity of these paths. For a hypercubic lattice with coordination number $z$, the summation over neighbors yields $\eta = 2z$. This factor of $2$ arises precisely because there are two distinct types of interfering pathways (Type A and Type B), each contributing $z$ valid intermediate states.

\subsection{Optimal Cancellation Condition}
Combining the results from Appendix~\ref{app:second_order} and the above derivation, the total effective pair hopping amplitude $\tilde{J}_{\text{eff}}$ up to the third order is
\begin{align}
    \tilde{J}_{\text{eff}} = J_p + \frac{2J^2}{U} - \eta \frac{J^2 J_p}{U^2}.
\end{align}
To suppress the transport, we set $\tilde{J}_{\text{eff}} = 0$. Solving for $J_p$ yields the optimal hopping strength:
\begin{align}
    J_p^{\text{opt}} = \frac{2J^2 U}{\eta J^2 - U^2}.
\end{align}
This expression justifies the optimal parameters $J_p^{\text{opt, 1D}}$ and $J_p^{\text{opt, 2D}}$ used in the numerical simulations in the main text.

\subsection{Generalization to Multi-Doublon Systems}

Although the derivation becomes analytically more complex compared to the single-doublon case, the presence of additional doublons does not fundamentally alter the underlying transition channels. Given that the essence of the SWT is to sum up the contributions from multi-order transition pathways, the conclusion derived for a single doublon can be naturally generalized to the multi-doublon scenario.

\begin{figure}[ht]
    \centering
    \includegraphics[]{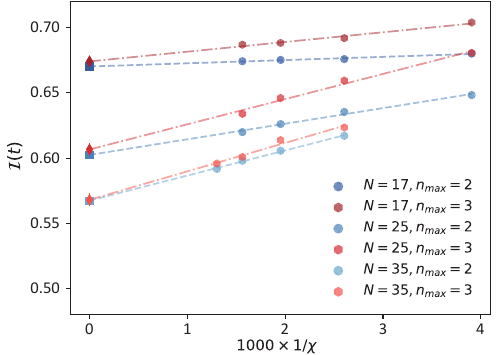}
    \caption{Validation of the maximum filling constraint. The imbalance $\mathcal{I}(t)$ is plotted as a function of the inverse bond dimension $1/\chi$ for representative system sizes $N=17$, $25$, and $35$, with the evaluation time and Hamiltonian parameters chosen identically to Fig.~\ref{fig:mb_fss}. Circle and hexagon markers denote the results obtained under $n_{\text{max}}=2$ and $n_{\text{max}}=3$, respectively, while square and diamond markers represent their corresponding extrapolated values in the infinite bond dimension limit, where all data for $n_{\text{max}}=2$ is identical to that in Fig.~\ref{fig:mb_fss}. Dashed lines indicate the linear fits. Error bars indicate the fitting uncertainty but are negligibly small and largely obscured by the markers.}
    \label{fig:max_filling}
\end{figure}

\section{Details of Numerical Simulations}
\label{app:sim_details}

To ensure the full reproducibility of our numerical results, we detail the explicit parameters and methodologies used in our simulations. In the strongly interacting BH model ($u = 10$) initialized with doublon configurations, the dynamical generation of triplons (three particles per site) or higher occupations is highly off-resonant. Specifically, the energy cost to form a triplon from a doublon and a single particle is $\Delta E \approx \left|2U\right| \gg J$, rendering such transitions extremely suppressed. Consequently, for all numerical simulations, we truncated the local Hilbert space to a maximum occupation of $n_{\text{max}}=2$ bosons per site. This physically justified truncation effectively captures the principal dynamical features of the system while significantly improving the computational efficiency of both the Krylov subspace projection method and the TDVP simulations.

Fig.~\ref{fig:max_filling} presents the TDVP simulation results for representative system sizes under a relaxed constraint of $n_{\text{max}}=3$. As shown, although minor differences exist at finite bond dimensions $\chi$, the density imbalance values extrapolated to the infinite bond dimension limit are nearly identical under both filling constraints. This explicitly verifies that our maximum filling restriction does not alter the core conclusions. Furthermore, the overall dynamical behavior observed under this parameter choice is highly robust throughout the strongly interacting regime ($u \gtrsim 10$), where the large energy gap $\Delta E \gg J$ ensures that transitions to triplons or higher occupations remain highly off-resonant. Conversely, as $u$ decreases and departs significantly from the strong-coupling limit, both the $n_{\text{max}}=2$ state-space truncation and the interference-induced prethermal phenomena face severe challenges and may eventually break down, as the reduced energy gap can no longer suppress higher-occupancy states or guarantee the validity of the SWT.

For the single-doublon dynamics and small many-body systems, the unitary evolution was implemented using Krylov subspace projection methods. Specifically, we utilized the \texttt{krylovsolve} function from the open-source Quantum Toolbox in Python (QuTiP) \cite{qutip5}. This approach efficiently approximates the action of the time-evolution operator by projecting the high-dimensional state vector onto a significantly smaller Krylov subspace generated via Lanczos iterations. We operated the solver using its default tolerance and subspace dimension parameters, which yield highly accurate unitary dynamics for sparse Hamiltonians.

Our finite-size scaling analysis for larger lattices (up to $N=41$) utilized the TDVP algorithm implemented via the \texttt{ITensor} software \cite{itensor} in Julia \cite{julia}. Throughout all sweeps, a strict singular value decomposition cutoff of $10^{-9}$ was enforced to minimize truncation errors and accurately capture quantum correlations. The evolution initially proceeded under the $2$-site TDVP scheme, permitting the tensor network bond dimension to adaptively expand in response to the dynamic growth of entanglement. Once this bond dimension reached a predefined maximum value $\chi$, the algorithm seamlessly transitioned to the computationally more efficient $1$-site TDVP scheme. Furthermore, the simulations  (using a time step $dt = 0.001 \times 2\pi/\left|J_{\text{eff}}\right|$) were performed using varying maximum bond dimensions $\chi$ structurally tailored to the respective system sizes: $\chi \in \{256, 384, 512, 640\}$ for $N=17, 21,$ and $25$; $\chi \in \{384, 512, 640, 768\}$ for $N=31$ and $35$; and $\chi \in \{512, 640, 768, 896\}$ for the largest system size $N=41$. This systematic scaling approach allows for a rigorous extrapolation of the density imbalance to the infinite bond dimension limit.

\begin{figure*}[ht]
    \centering
    \includegraphics[]{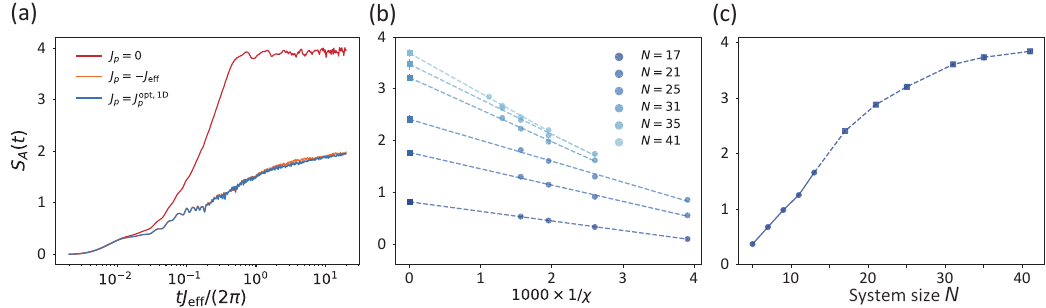}
    \caption{Finite-size scaling of the long-time bipartite entanglement entropy. (a) Time evolution of the bipartite entanglement entropy $S_A\left(t\right)$ for $N=13$. The red, orange, and blue curves correspond to $J_p=0$, $-J_{\text{eff}}$, and $J_p^{\text{opt, 1D}}$, respectively. (b) The bipartite entanglement entropy evaluated at $t=2\times2\pi/|J_{\text{eff}}|$ for larger system sizes ($N>13$), calculated using the TDVP algorithm with various bond dimensions $\chi$. The dashed lines represent linear fits used for extrapolation to the infinite bond dimension limit, with the predicted values denoted by square markers. Error bars indicate the fitting uncertainty but are negligibly small and largely obscured by the markers.  (c) The bipartite entanglement entropy at $t=2\times2\pi/|J_{\text{eff}}|$ plotted as a function of the system size $N$. Circular markers represent exact numerical results obtained via the Krylov subspace projection methods for smaller systems, while square markers represent the TDVP-extrapolated predictions, corresponding directly to the square markers in panel (b).}
    \label{fig:entropy}
\end{figure*}

\section{Supplementary Evidence of Prethermalization}
\label{app:entropy}

To further substantiate the prethermalization scenario observed via the imbalance in the main text, we analyze the dynamics of the bipartite entanglement entropy $S_A\left(t\right)$. We divide the 1D chain of $N$ sites into two spatial subsystems, $A$ and $B$, where subsystem $A$ encompasses the first $N_A = \lfloor N/2 \rfloor$ sites. The bipartite entanglement entropy is defined as the von-Neumann entropy of the reduced density matrix \cite{bennett1996concentrating, calabrese2005evolution, amico2008entanglement}:
\begin{align}
    S_A(t) = -\text{Tr}\left[\hat{\rho}_A(t) \ln \hat{\rho}_A\left(t\right)\right],
\end{align}
where $\hat{\rho}_A(t) = \text{Tr}_B\left[\hat{\rho}\left(t\right)\right]$ is obtained by tracing out the degrees of freedom in subsystem $B$.

As illustrated in Fig.~\ref{fig:entropy}(a), for a small system size ($N=13$), the uncontrolled system ($J_p=0$) exhibits a rapid growth of $S_A(t)$ that quickly saturates to a high value, signaling fast thermalization. In stark contrast, under the optimal pair-hopping condition ($J_p = J_p^{\text{opt, 1D}}$), the entropy exhibits a remarkably slow, logarithmic growth following an initial rapid increase. This behavior aligns with the characteristic entanglement dynamics of a prethermal phase.

To conclusively rule out finite-size effects and confirm that this suppressed entanglement growth persists at macroscopic scales, we also perform a finite-size scaling analysis using the TDVP algorithm for systems up to $N=41$. Similar to the imbalance analysis, Fig.~\ref{fig:entropy}(b) shows the entropy evaluated at a representative time ($t = 2 \times 2\pi/\left|J_{\text{eff}}\right|$) for various bond dimensions $\chi$, rigorously extrapolated to the infinite bond dimension limit. Fig.~\ref{fig:entropy}(c) presents this extrapolated entropy as a function of the system size $N$. The systematic scaling curve conclusively demonstrates that the significant suppression of entanglement spreading remains fundamentally robust in the macroscopic limit. 

\bibliography{references}

\end{document}